\documentclass[epj]{svjour}
\usepackage{graphics}
\usepackage{graphicx} 
\usepackage{lipsum}
\usepackage{color}
\usepackage{mathtools}
\usepackage{cuted}
\newcommand{\R}{}

\newcommand{\thex}{\ensuremath{\theta_1}}
\newcommand{\thel}{\ensuremath{\theta_2}}
\newcommand{\Br}{\ensuremath{\mathcal{B}}}
\newcommand{\Jm}{\ensuremath{{J}^{\mathrm{max}}}}
\newcommand{\lag}{\ensuremath{\epsilon_\gamma}}
\newcommand{\all}{\ensuremath{\alpha_\Lambda}}
\newcommand{\cox}{\ensuremath{\cos{\theta_X}}}

\newcommand{\acox}{\ensuremath{\left|\cos(\theta_X)\right|}}
\newcommand{\lp}{\ensuremath{\Lambda\bar{p}}}

\newcommand{\Btwo}{Belle~II\,}
\begin{document}
\title{\boldmath On measurement of photon polarization in radiative 
penguin $B$ decays to baryons}
\author{P.~Pakhlov\inst{1}\inst{2} \and T.~Uglov\inst{1}
}                     
%
%
\institute{P.~N.~Lebedev Physical Institute of the RAS, Moscow, Russia\and Higher School of Economics,  Russian Federation}
%
%
\abstract{
A measurement of the photon polarization in radiative penguin $B$ decays 
provides a test of the Standard Model and a probe for New Physics, that can 
lead to a deviation from the Standard Model prediction of left-handed photons 
in $b\to s \gamma$. We propose a new method to measure the photon polarization 
using the baryonic decay $B^- \to \Lambda\bar{p} \gamma$. The $P$-violating 
$\Lambda$-hyperon decay allows a measurement of the $\Lambda$ helicity to be 
performed, which can be uniquely related to the photon polarization in a 
model-independent way. The $B^- \to \Lambda\bar{p} \gamma$ decay was recently 
measured to have a large branching fraction providing a possibility to get 
meaningful results with the data already available at LHC and B-factory 
experiments. An increase of the $B$-meson sample at high luminosity LHC 
experiments and \Btwo should provide a really stringent test by using this 
method already in the near future. 
} 
\maketitle

Due to the chiral structure of the $W$-boson interaction with fermions in the Standard Model (SM), the photon polarization in penguin $b \to s \gamma$ transition reveals a maximal parity violation: the emitted photon is polarized left-handed in the decays of $B$ mesons {\R (right-handed in $\overline{B}$ decays). While most theoreticians agree that admixture of ``wrong'' polarization is tiny in the SM, of the order of $m_s /m_b \sim 0.02$~\cite{SM}, some of them suppose the hadronic effects can enhance the right-handed photons in exclusive channels up to $\Lambda_{\mathrm{QCD}} / m_b \sim 0.05$ and inclusively up to $g_s/(4\pi)$~\cite{Grinstein:2004uu}. In any case, the SM prediction of photon polarization in $b \to s \gamma$ remains quite accurate to serve as the SM test.} 


In many SM extensions the right-handed contribution can be enhanced. 
Supersymmetric scenarios suggest several such mechanisms, {\it e.g.} 
an intermediate charged Higgs gives rise to the $b_R \to s_L$ transition in 
the models with $R$-parity violation~\cite{R-parity}, while 
in unconstrained MSSM the chirality flip along the gluino line in the loop 
involving left-right squark mixing can inverse the SM prediction, resulting 
in right-handed photons~\cite{Gluino}. In the left-right symmetric model 
chirality flip along the $t$-quark line in the loop involves $W_L - W_R$ 
mixing~\cite{LR}. It was found that in certain allowed regions of the 
parameter space of all these models photons emitted in $b \to s \gamma$ can 
be largely right-handed polarized, without affecting the SM 
prediction for the inclusive radiative decay rate. 

A clear SM prediction and a possibility to check models that have not yet been 
refuted by numerous tests motivate a precise measurement of the photon 
polarization. The only problem in performing such a test is a lack of practical
methods to access it experimentally. Very few methods were proposed so far: 
one is to use $B^+ \to K^+ \pi^+ \pi^- \gamma$ decays, where an angular 
analysis of the photon direction with respect to the $K^+ \pi^+ \pi^-$ plane 
in their center-of-mass frame allows to distinguish between differently 
polarized photons through the measurement of the up-down asymmetry, which is 
related to the photon polarization 
\lag~\cite{Gronau:2001ng,Kou:2010kn,Gronau:2017kyq}. The experimental 
probe with this method performed by LHCb~\cite{Aaij:2014wgo} has demonstrated 
non-zero photon polarization with a high significance. However, no quantitative
result on the degree of polarization was obtained. Another possibility to 
probe for a right-handed photon component from $b \to s \gamma$ is to search 
for indirect CP violation in $B^0_{(s)} \to X^0 \gamma$, which can only arise 
if the ``wrong'' component of the photon polarization allows 
$B^0_{(s)}$ and $\overline{B}{}^0_{(s)}$ interference~\cite{Atwood:1997zr,Muheim:2008vu}.
A recent LHCb analysis~\cite{Aaij:2016ofv} showed no significant 
CP violation, confirming that only one polarization dominates, but again 
without a quantitative limit. {\R The third approach is based on angular analysis of the $B\to K^* e^+e^-$ decay, where $e^+e^-$ pair originates from the virtual photon~\cite{Grossman:2000rk}. This study limited to the region of very low $q^2$ ($e^+e^-$ pair invariant mass) was performed by LHCb collaboration \cite{LHCb:2015ycz,LHCb:2020dof} setting so far the best bounds on the virtual photon polarization.}
 
In this paper we propose a new promising method to measure directly the photon 
polarization using exclusive radiative $B^-$ decays to $\Lambda \bar{p} \gamma$. 
This final state is unique due to possibility to measure the polarization of 
$\Lambda$ through the $P$-violating asymmetry in its decay, {\it e.g.} into $p \pi^-$. 
While in general the $\Lambda$ polarization does not always correspond to photon 
polarization, there is a region in the decay phase space, where it is possible to relate 
two polarizations in a model-independent way. This decay was first observed by 
Belle~\cite{Lee:2005fba} with a relatively large branching fraction and a high signal 
purity, thus suitable for such studies.

\begin{figure}[htb]
\centering 
\includegraphics[width=0.48\textwidth]{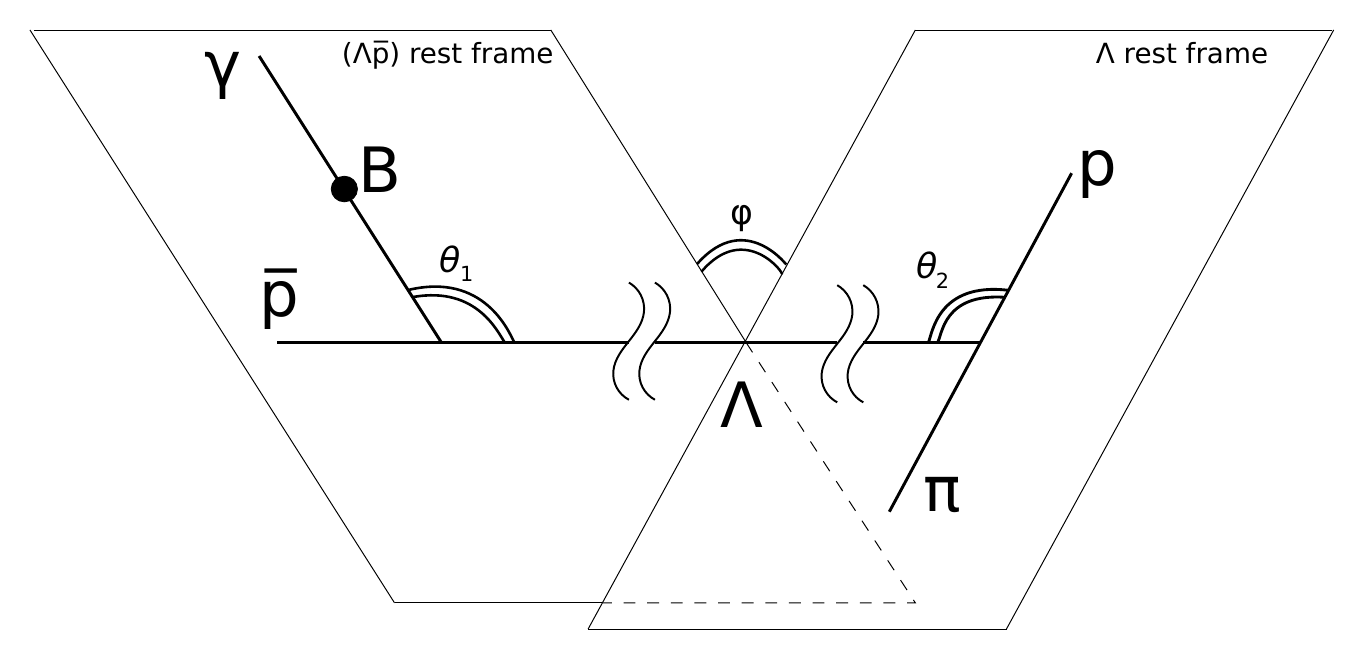}
\caption{\label{fig:kinematics} Kinematics of the $B^- \to \Lambda \bar{p} \gamma,\,\Lambda\to p \pi^-$ decay.}
\end{figure}

Let us first define the observables available in the experiment for the photon
polarization studies in the proposed decay chain $B^- \to \Lambda \bar{p} \gamma$, followed by
$\Lambda \to p \pi^-$. For the sake of simplicity we denote the $(\Lambda \bar{p})$ combination 
by $X$ throughout the paper. This does not imply that we consider $X$ as a 
real state or having definite quantum numbers. The final state is characterized
by the $X$ mass and three angles (see Fig.~\ref{fig:kinematics}):
the $X$ decay angle (the angle between photon- and
$\Lambda$-momentum directions in the $X$ rest frame), \thex; the $\Lambda$ 
decays angle (the angle between antiproton- and proton-momentum directions in 
the $\Lambda$ frame), \thel; and the angle between $B$ and $\Lambda$ decay 
planes ({\R the angle between vector products of $\vec{p}_\gamma\times\vec{p}_\Lambda$ and $\vec{p}_p\times\vec{p}_\pi$ in the $X$ rest frame}), $\varphi$. We 
demonstrate that the photon polarization, defined as
$\lag \equiv \frac{\Br_L-\Br_R}{\Br_L+\Br_R}$,  where $\Br_{L(R)}$ are the 
$B$-decay branching fractions to the photon of helicity $-1(+1)$, can be 
related to the angular variables and thus extracted from the experimental fit.

The system $X$ likely represents the $S$-wave $(\lp)$ combination (thus having 
$J^P=1^-$ quantum numbers), as the Belle studies~\cite{Lee:2005fba,Wang:2007as}
demonstrated a strong $(\lp)$ near-threshold enhancement while 
the larger waves should be suppressed near the threshold. This is supported 
by the measured $X$ angular distribution consistent with being flat and the 
fact that the observed $X$-mass shape corresponds to the off-shell 
$K^{*-} \to \Lambda \bar{p}$ decay. Under assumption of only $(\lp)$ in the 
$S$-wave contribution, the angular part of the decay matrix element has a 
very simple form
\begin{equation}
\mathcal{S} \sim 1 + \lag \all \cos{\thex} \cdot \cos{\thel},
\label{kstar}
\end{equation}
where $\alpha_\Lambda$ -- is the $P$-violating parameter in the 
$\Lambda \to p \pi^-$ decay.

If the dominance of the $S$-wave could be proved theoretically, this simple 
formula would serve to fit the data, since the parameter $\alpha_\Lambda$  
is known here, and the only free fitting parameter is the unknown photon 
polarization. However, it is hardly possible today to rigorously prove using 
theoretical arguments or calculations, that  the  $\Lambda \overline{p}$ 
production dynamics is really dominated by the S-wave contribution only.  
Thus, there are two options to accomplish this task: either to find 
kinematic regions free of model dependencies, or to experimentally estimate 
the dynamics and constraints on the model uncertainty of the measured value 
directly from the data. 

It turns out that the first option can be implemented. There is a kinematic 
region, where the relationship between the photon and $\Lambda$ polarizations 
can be described unambiguously without relying on the model assumptions. 
This is the case of $\cox = \pm 1$, that is, when $\Lambda$ moves exactly 
forward or backward in the $\Lambda p$ rest frame, thus the momenta of all 
three daughter particles $\gamma,~\Lambda,~\bar{p}$ are collinear. 
Indeed, whatever the angular momenta in the $B \to \gamma X$ and $X \to \lp$ 
decays, their projections on the common decay axis are equal to 0. Hence the 
sum of the projections of the spins $\Lambda$ and $\bar{p}$ on this axis is 
exactly equal to the projection of the photon spin. While the proton spin 
direction is unmeasurable, the photon transversity ensures that its spin 
projection to this axis is $\pm1$, which should be equal to the sum of two 
one halves, thus the proton spin coincides with the spin of $\Lambda$.

Although the phase space of this kinematic region is vanishing, experimentally 
it is  possible to extrapolate to the region 
$\left| \cos (\theta_X) \right| = \pm1$ by studying $\lambda_\Lambda(\cox)$ in 
the vicinity of $\cox=\pm1$. Technically, this will require a fit of the 
angular distribution of $\cos(\theta_\Lambda)$ with the function 
$1 + A \cos{\theta_\Lambda} + B \sin{\theta_\Lambda}$ in bins of \cox. 
If one draws then the dependence of the measured $A(\cox)$ and $B(\cox)$ and 
extrapolate them into $\acox=1$, then the value of 
$A_\Lambda=\mathrm{sign}(\cox)A|_{\cox=\pm1}$ can be related to the proton 
polarization via $\lag=\frac{A_\Lambda}{\all}$, while $B|_{\cox=\pm1}$ should be 0.
 
A more mathematically rigorous approach can be applied, but it requires a 
simple model assumption, namely, a constraint on the maximum total orbital 
angular momentum of the system $X$: to fit the angular distributions with a 
general function, where the QCD dynamics is represented by free parameters. 
This approach has the advantage that it uses the entire kinematic decay region,
rather than its small part, and the output will not only give a measurement 
of the photon polarization, but also the parameters describing the QCD decay 
dynamics, which are of additional interest. The angular part of the matrix 
elements for the studied process is convenient to be written in the helicity 
formalism
as follows:
\begin{strip}
\begin{equation}
\mathcal{S}= \! \! \!
\mathop{\sum_{\lambda_\gamma=\pm1}}_{\lambda_p,\lambda_{\bar{p}}=\pm1/2} \bigg|
\mathop{\sum_{J_X=1\dots \Jm} } \, \,  \mathop{\sum_{\lambda_X=-J_X\dots J_X}}_{\lambda_\Lambda=\pm1/2}
\left( D^0_{0,\lambda_X-\lambda_\gamma}  
\cdot a_{\lambda_X,\lambda_\gamma}\right)
\cdot
\left(D^{J_X}_{\lambda_X,\lambda_\Lambda-\lambda_{\bar{p}}}  
\cdot b^{J_X}_{\lambda_\Lambda,\lambda_{\bar{p}}}\right)
\cdot
\left(D^{1/2}_{\lambda_\Lambda,\lambda_{p}-\lambda_{\pi}}
\cdot c_{\lambda_{p},\lambda_{\pi}}\right)
\bigg|^2,
\label{amplitude:1}
\end{equation}
\end{strip}

\noindent where the first, second and third Wigner $D$-functions describe $B$ decay
to photon and $X$, the $X$ transition to $\Lambda$ and antiproton, and 
$\Lambda\to p \pi^-$ decay, respectively. Complex constants $a$, $b$ and $c$ 
define contributions of various helicity states to the 
total decay amplitude. The value of the $X$ total orbital momentum, $J_X$, 
should be limited by some number, \Jm\, to avoid infinite numbers of 
parameters, while the minimal value is equal to 1, as 0 is forbidden by 
momentum conservation. Since the values of $J_X$ and the polarization of 
the $\Lambda$ baryon in contrast to the polarization of the photon, proton and 
antiproton, can not be measured, corresponding amplitudes must be summed 
coherently before squared.

Taking into account that pions and $B$ mesons are scalars and that 
$D^0_{0,\lambda_X-\lambda_\gamma}$ not vanishes only if $\lambda_X=\lambda_\gamma$, 
one can simplify the equation~(\ref{amplitude:1}) to:
\begin{multline}\label{amplitude:2}
S \sim \! \! \! \! \! \! \! \mathop{\sum_{\lambda_\gamma=\pm1}}_{\lambda_p,\lambda_{\bar{p}}=\pm1/2} 
 \bigg|  a_{\lambda_\gamma} 
\mathop{\sum_{J_X=1\dots\Jm}}_{\lambda_\Lambda=\pm1/2} 
\left( D^{J_X}_{\lambda_\gamma,\lambda_\Lambda-\lambda_{\bar{p}}}
 \cdot b^{J_X}_{\lambda_\Lambda,\lambda_{\bar{p}}}  \right)
\cdot \\
\left(
D^{1/2}_{\lambda_\Lambda,\lambda_{p}}
\cdot c_{\lambda_{p}}\right)
\bigg|^2.
\end{multline}

The matrix element (\ref{amplitude:2}) depends on $4\times J_X$ complex 
amplitudes $b^{J_X}_{\pm1/2,\pm1/2}$, and the other two amplitudes $c_{\pm1/2}$. 
The magnitudes of the latter two are, in turn, known from the measured 
$P$ asymmetry in $\Lambda$ decays~\cite{pdg}:
\begin{equation}
\alpha_\Lambda=   \frac{|c_{+1/2}|^2-|c_{-1/2}|^2}{|c_{+1/2}|^2+|c_{-1/2}|^2}=0.732\pm 0.014.
\label{lambda_decay_assymetry}
\end{equation}
The phases of the $c$ are unobservable, and hence cancel out.

The expressions (\ref{amplitude:2})-(\ref{lambda_decay_assymetry}) explicitly 
define the ready-made function that can be directly used in the fit to the 
angular variables in the experimental data to extract $\lag$, which in this 
parameterization is defined as $\lag =\frac{a^2_{-1}-a^2_{+1}}{a^2_{-1}+a^2_{+1}}$.
The number of free parameters in addition to the sought \lag\ depends on the 
assumption about the maximum orbital momentum of the $X$ system. Up to now only 
$K^{*(*)}$ with spin 1 or 2 were observed in the $B$ radiative 
decays~\cite{pdg}, thus it is justified to assume $\Jm=2$. Moreover, high 
$J_X$ suppression by relativistic phase space (much reduced in case of the 
baryon-antibaryon final state and low $X$ mass tendency) supports this 
assumption. Limiting $\Jm \leq 2$ there are seven complex variables in 
addition to the overall normalization and \lag\ that describe  
$X \to \Lambda \bar{p}$ dynamics in the fit function. 

\begin{figure}[htb]
\centering 
\includegraphics[width=0.5\textwidth]{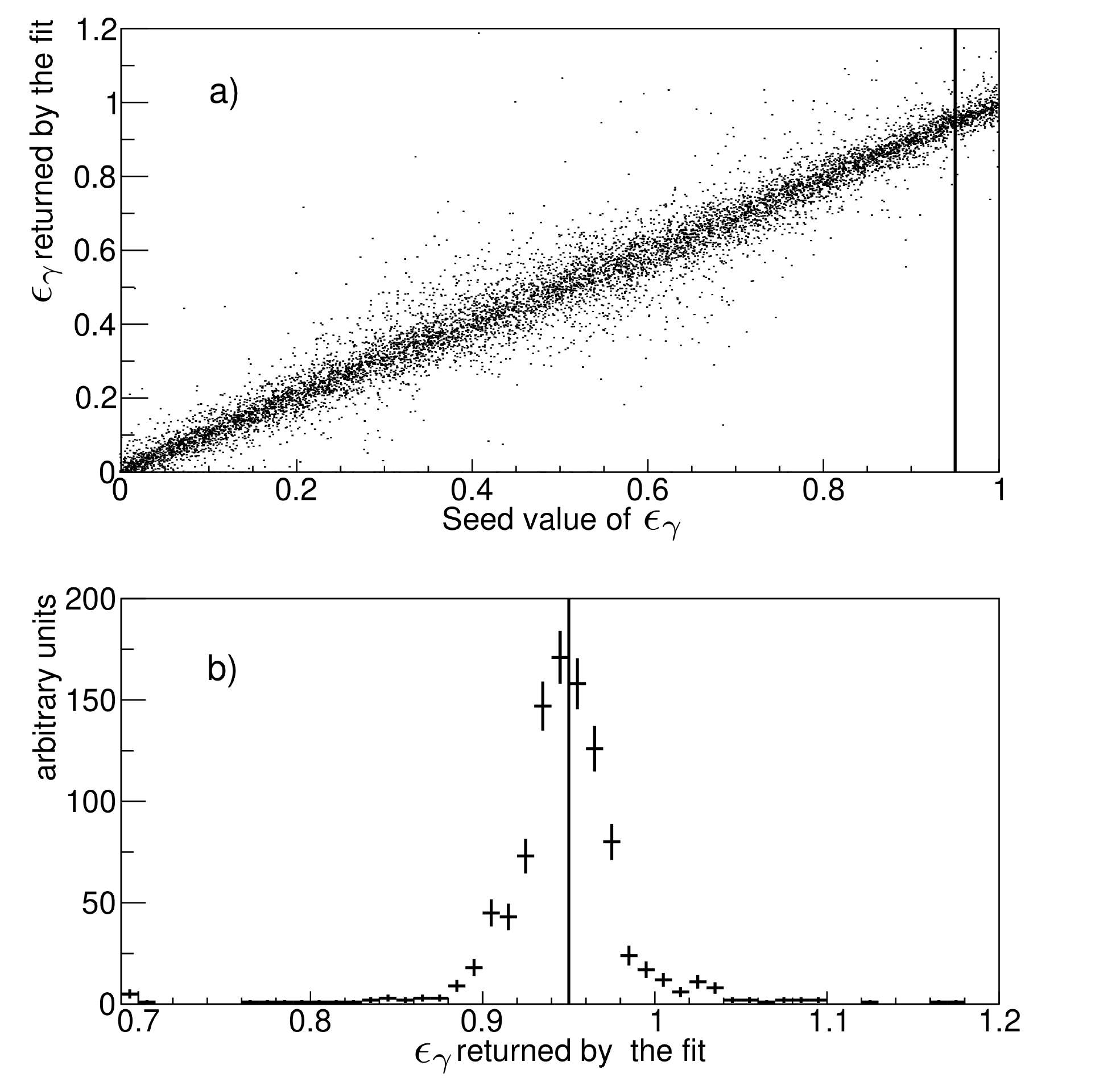}
\caption{\label{fig:res}  a) Distribution of \lag, returned by fits to the 
Monte Carlo samples, depending on the seed value. b) \lag\ spectrum with 
the \lag\ seed value fixed to 0.95. Vertical lines show 
${\epsilon_{\gamma}}_{\mathrm{seed}}=0.95$.}
\end{figure}

We check the capability of the near future experiments 
(\Btwo~\cite{Abe:2010gxa} and LHC experiments under high luminosity) to 
extract \lag. It is easy to estimate that \Btwo will be able to reconstruct 
approximately 8000 signal events with its full data set, corresponding to the 
$50~\mathrm{ab}^{-1}$ integrated luminosity. We generate 10000 samples with 
random parameters $b^{J_X}_{\pm1/2,\pm1/2}$ using toy Monte Carlo simulation. 
The  parameters $b^{J_X}_{\pm1/2,\pm1/2}$  for $J_X=1,2$ are uniformly distributed 
within the circle of radius 10, thus representing all possible models of the 
studied decay. Each sample consists of 8000 ``signal'' events generated 
according to (\ref{amplitude:2}). For all samples we perform an unbinned 
likelihood fit with the fitting function defined as:
\begin{equation}
PDF(\thex,~\thel,~ \varphi) \propto \lag \mathcal{S}_{-1}+(1-\lag) \mathcal{S}_{+1} \, ,
\label{fitfunction}
\end{equation}
where $\mathcal{S}_{\lambda_\gamma=\mp1}$ are normalized matrix elements for two 
photon polarizations (\ref{amplitude:2}) that include all $(b^{J_X}_{\pm1/2,\pm1/2})$ as free parameters. It was found that almost all fits converged during 
automatic fitting, and the procedure is able to extract the true (generated) 
values of helicity parameters, while the desired result of \lag\ extraction is 
unbiased by the fit. The scatter plot of generated and reconstructed \lag\ 
values shown  in Fig.~\ref{fig:res}~a) demonstrates linear dependence and 
scattering compatible with the expected statistical fluctuations. In 
Fig.~\ref{fig:res}~b) the spread of the returned values at ${\epsilon_{\gamma}}_{\mathrm{seed}}=0.95$ is shown. The mean value of the resulting ${\epsilon_{\gamma}}$ distribution is $0.948\pm 0.013$ which is compatible with the seed value of 
$0.95$ while the width is $0.02$ and is described by the fit error. 

The presently available data is likely insufficient for both model-independent 
methods due to a large number of free parameters in the fit. Belle has doubled 
the data sample since the last publication of the 
$B^-\to \Lambda \bar{p}\gamma$ measurements, and can thus reconstruct up to 200 
signal events. Though LHCb has a higher production rate, due to its much 
smaller $\Lambda$ reconstruction efficiency, the number of reconstructed 
events at full current integrated  LHCb luminosity can be estimated 
comparable to the Belle one~\cite{Aaij:2016xfa}. While use of the 
model-independent methods is not feasible now both at Belle/BaBar and LHCb, 
the meaningful measurements can still be done. The main concern of 
the model-dependent study is to keep the model uncertainties under control. 

We thus propose to use the assumption about the $S$-wave and perform a fit 
to the data using the formula~\ref{kstar}. The main task will be to estimate 
the model uncertainties. This can be done at the level of existing statistical 
errors if the same data is used to restrict the parameters of the models. It 
is necessary to scan all models, discard those of them that contradict the 
angular distribution of the data (for example, by 90\% CL), and of the 
remaining ones, check the spread of the extracted parameter \lag\ from the 
generated one. We estimated that with 200 signal events, with a 
signal-to-background ratio of $1:1$, one can expect a statistical accuracy 
of \lag\ around 0.4 (while the value itself can vary from $-1$ to $+1$) using 
toy MC. Although the model uncertainties depend on how well the data will 
match the $S$-wave distribution, we encourage to perform such an analysis for 
the first measurement of the \lag.

{\R 
Due to low statistic of the currently available data samples, it could also be useful to check, whether \lag\ is zero or not. Belle attempted to search for $P$-parity violation in $B^-\to \Lambda \bar{p}\gamma$ decay by study of $\Lambda$ polarization~\cite{Wang:2007as}. However, the answer obtained with this method is not informative, as $\Lambda$ polarization could vanish after integration over all $\cos(\theta_X)$ region even in presence of $P$ violation. The correct approach is to search for the correlation between $\cos(\theta_X)$ and $\Lambda$ polarization, which is obviously absent in case of $\lag=0$. A presence of significant correlation would prove non-zero photon polarization without any assumption on decay dynamics. As a formal correlation function covariation $cov(\cos(\theta_X), \cos(\theta_\Lambda)) = \sum \cos(\theta_X) \cdot \cos(\theta_\Lambda)$ can be used. We estimate the ability of Belle to exclude zero photon polarization using toy Monte Carlo simulation. For 200 events, corresponding to the current Belle statistics and generated in $S$-wave assumption, the mean covariation value is $0.081$ and its distribution has a width of $0.028$, which means that $\sim 3\sigma$ indication for non-zero photon polarization can be obtained by Belle.}


In summary, we conclude that the measurement of the photon polarization in 
rare radiative B decays, which is very sensitive to possible NP contributions, 
is feasible using $B^-\to\Lambda\bar{p}\gamma$ decay. The access to the photon 
polarization is provided by a measurement of the $\Lambda$ polarization. 
We propose a method which allows one to accurately measure the polarization 
of a photon in a model-independent way using the large statistics of the 
near-future $B$ experiments. We also checked that it is possible to perform 
the first measurement of the polarization of a photon using the existing 
relatively small statistics, while the model dependence of the measurement will 
remain under control. 

The work of P.~Pakhlov was conducted within the framework of the Basic 
Research Program at the National Research University Higher School of 
Economics (HSE). T.~Uglov acknowledges the support of the Russian Ministry 
of Science and Higher Education under contract\\ 14.W03.31.0026.

%
%
%
%
%

\end{document}